\documentclass{sig-alternate-05-2015}
\usepackage{color}
\usepackage{dsfont}
\usepackage{amsmath}
\usepackage{subfig}
\usepackage{graphicx}
\usepackage{multirow}
\usepackage[table]{xcolor}
\usepackage{pgfplots}
\usepackage[]{todonotes}
\usepackage[named]{algo}
\usepackage{enumitem}
\usepackage[switch]{lineno}
\DeclareMathOperator*{\argmax}{arg\,max}

\newcommand{\figwidth}{0.47\columnwidth}
\makeatletter
\renewcommand{\paragraph}{\@startsection{paragraph}{4}{\z@}%
  {-3.25ex\@plus -1ex \@minus -.2ex}%
  {1.5ex \@plus .2ex}%
  {\normalfont\normalsize\bfseries}}
\makeatother

\begin{document}
\setlength{\abovedisplayskip}{3pt}
\setlength{\belowdisplayskip}{3pt}





%

\CopyrightYear{2016} 
\setcopyright{acmcopyright}
\conferenceinfo{CIKM'16 ,}{October 24-28, 2016, Indianapolis, IN, USA}
\isbn{978-1-4503-4073-1/16/10}\acmPrice{\$15.00}
\doi{http://dx.doi.org/10.1145/2983323.2983895}


\clubpenalty=10000 
\widowpenalty = 10000

\title{Adaptive Distributional Extensions to DFR Ranking}
%
%
%
%
%

\numberofauthors{4} 
%
\author{
\alignauthor
Casper Petersen\\
       \affaddr{University of Copenhagen}\\
       \email{cazz@di.ku.dk}
\alignauthor
Jakob Grue Simonsen\\
       \affaddr{University of Copenhagen}\\
       \email{simonsen@di.ku.dk}
\and
\alignauthor
Kalervo J\"{a}rvelin\\
       \affaddr{University of Tampere}\\
       \email{Kalervo.Jarvelin@staff.uta.fi}
\alignauthor 
Christina Lioma\\
       \affaddr{University of Copenhagen}\\
       \email{c.lioma@di.ku.dk}
}

\maketitle
\begin{abstract}
Divergence From Randomness (DFR) ranking models assume that informative terms are distributed in a corpus differently than non-informative terms. Different statistical models (e.g.\ Poisson, geometric) are used to model the distribution of non-informative terms, producing different DFR models. An informative term is then detected by measuring the divergence of its distribution from the distribution of non-informative terms. However, there is little empirical evidence that the distributions of non-informative terms used in DFR actually fit current datasets. Practically this risks providing a poor separation between informative and non-informative terms, thus compromising the discriminative power of the ranking model. We present a novel extension to DFR, which first detects the best-fitting distribution of non-informative terms in a collection, and then adapts the ranking computation to this best-fitting distribution. We call this model Adaptive Distributional Ranking (ADR) because it adapts the ranking to the statistics of the specific dataset being processed each time. Experiments on TREC data show ADR to outperform DFR models (and their extensions) and be comparable in performance to a query likelihood language model (LM).

\end{abstract}

%
%


%
%

%
%



\section{Introduction}
Early work on automatic indexing \cite{2poisson,harterp1,harterp2} observed that 
\textit{informative} words, e.g. those belonging to a technical vocabulary, are distributed in a document collection differently than \textit{non-informative} words, e.g. those usually treated as stopwords. Specifically, the difference in the distributions of these two different types of words was that informative words were observed to appear more densely in few so-called \textit{elite} documents. Based on this, non-informative words were modelled by a Poisson distribution, and it was hypothesised that an informative word can be detected by measuring the extent to which its distribution deviates from a Poisson distribution, or, in other words, by testing the hypothesis that the word distribution on the whole document collection does not fit the Poisson model \cite{Amati:2002:PMI:582415.582416}. This so-called 2-Poisson model has led to well-known ranking models such as BM25 \cite{Robertson1994} or Divergence from Randomness (DFR) \cite{Amati:2002:PMI:582415.582416,terrierDFR}.



For DFR ranking models in particular (presented in Section \ref{s:dfr}), the idea that non-informative terms tend to follow a specific distribution is central. 
Different distributions (e.g.\ the Poisson or geometric distribution) can be used to obtain different DFR ranking models. The goal is to choose a distribution that provides a good fit to the empirical distribution of 
non-informative terms in $C$. However, there is little empirical evidence that the distributions used in DFR actually fit current datasets. Practically this risks providing a poor separation between informative and non-informative terms, thus compromising the discriminative power of the ranking model.



Motivated by this, we present a novel extension to DFR, which first detects the best-fitting distribution (among a candidate set of distributions) of non-informative terms in a collection, and then adapts the ranking computation to this ``best-fitting'' distribution. We call this model Adaptive Distributional Ranking (ADR, presented in Section \ref{s:adr}) because it adapts the ranking to the statistics of the specific dataset being processed each time. 
With our approach, only one ranking model is obtained per collection: the one that provides the best fit to the distribution of non-informative terms in that collection. 
Evaluation on TREC datasets (Section \ref{s:eval}) shows that our ADR models outperform the original DFR models, as well as more recent DFR extensions and perform on a par with a query likelihood LM \cite{Ponte:1998:LMA:290941.291008}.

\section{Divergence From Randomness}\label{s:dfr}
Given a query $q$ and a document $d$, a DFR ranking model estimates the relevance $R(q,d)$ of $d$ to $q$ as \cite{Amati:2002:PMI:582415.582416}:
\begin{equation}\label{ch4:eqn:DFR}
R(q,d) = \textstyle\sum_{t\in q} f_{t,q}\cdot\left(-\log_2 P_1\right)\cdot\left(1 - P_2\right)
\end{equation}
where $t$ is a term, $f_{t,q}$ is the frequency of $t$ in $q$, $-\log_2 P_1$ is the information content of $t$ in $d$, and $(1 - P_2)$ is the risk of accepting $t$ as a descriptor of $d$'s topic. 
We explain each of these two components next.

The information content $P_1$ measures the divergence of $f_{t,d}$ from $f_{t,C}$, where a higher divergence means more information is carried by $t$ in $d$. The assumption is that terms that bring little information are distributed over all documents in a certain way across the entire corpus; different distributions (called ``models of randomness'' in DFR) give rise
to different DFR ranking models. For example, using a Poisson distribution, $P_1$ is:
\begin{equation}\label{eqn:dfr_poisson}
P_1(t,\lambda,d) = e^{-\lambda}\lambda^{\hat{f}_{t,d}}/\hat{f}_{t,d}!
\end{equation}
where $\lambda\!=\!f_{t,C}/\vert C\vert$ is the parameter of the Poisson distribution, and $\hat{f}_{t,d}\!=\!f_{t,d}\cdot\log_2\!\left(1\! +\! c\cdot\mbox{avg\_l}/\vert d\vert\right)$ is a logarithmic term frequency normalisation where $c$ is a free parameter, avg\_l is the average document length in the collection, and $\vert d\vert$ is the length of $d$ (the number of terms). 
Eqn.\ \ref{eqn:dfr_poisson} returns the probability of seeing $\hat{f}_{t,d}$ occurrences of $t$ in $d$ and effectively tests whether $t$'s distribution on $C$ fits the Poisson distribution. If the probability of obtaining $\hat{f}_{t,d}$ occurrences of $t$  is low, then $t$ carries a high amount of information \cite{Amati:2002:PMI:582415.582416}. In addition to the Poisson distribution used in this example, DFR models can be instantiated also with the geometric, tf-idf (I$_n$), tf-itf (I$_F$) and tf-expected-idf (I$^e_n$) \cite{Amati:2002:PMI:582415.582416}.


The second component of Eqn.\ \ref{ch4:eqn:DFR}, the information gain, $P_2$, is the conditional probability of encountering $f_{t,d}\!+\!1$ occurrences of $t$ in $d$. 
If $P_2$ is high, the risk $(1 - P_2)$ associated with accepting $t$ as a descriptor of $d$ is low. Such ``after-effect'' sampling was taken \cite{Amati:2002:PMI:582415.582416} to be normalised either with Laplace: $P_2 = f_{t,d}/(f_{{t,d}}\!+\!1)$ 
or Bernoulli normalisation: $P_2 = (f_{t,C}\! +\! 1)/{\left(n_t\!\cdot f_{t,d}\! +\! 1\right)}$ 
where $n_t$ is the number of documents where $t$ occurs. 


\section{Adaptive Distributional Ranking}
\label{s:adr}
Our ADR models follow the basic DFR rationale presented above, but adapt the computation of $P_1$ in Eqn.\ \ref{ch4:eqn:DFR} to the best-fitting distribution of non-informative term collection frequencies for each collection. The ADR algorithm is shown below. 
\begin{algorithm}{ADR}[C]{
\qprocedure[C]{ADR}
\qinput{\!Collection $C$;\! $\mathbb{Y}\!=\!\{\mbox{parameterised statistical models}\}$}
}
\qfor the set $T$ of non-informative terms in $C$ \\
	\qfor each model $\mathcal{M}  \in \mathbb{Y}$ \\
   		find $\mathcal{M} $'s best parameters for fitting $T$
	\qendfor \\
	find $\hat{\mathcal{M} }$ that fits $T$ best\\
	replace $P_1$ in Eqn.\ \ref{ch4:eqn:DFR} by $\hat{\mathcal{M}}$
\qendfor \\
\qreturn an ADR model derived for $C$
\end{algorithm}

Given some collection $C$ and a set $\mathbb{Y}$ of candidate parameterised statistical models, we determine for each candidate model the optimal parameter values (if any) that make it fit how non-informative terms are distributed in $C$ (step 3 in the algorithm), select the best-fitting distribution (step 4), and plug it in the place of $P_1$ in Eqn.\ \ref{ch4:eqn:DFR} (step 5). The output is an ADR model adapted specifically to $C$ (i.e.\ a collection-specific ranking model). In this work, we use as candidate parameterised statistical distributions all the discrete models in \cite{psltois2015}, namely the geometric, negative Binomial, Poisson, power law and Yule--Simon. 
However, any other set of candidate distributions can be used in step 2. We explain steps 3 -- 5 next.

\subsection{Step 3. Optimal parameter fitting}
A parameterised statistical distribution $\mathcal{M}  = \left\{g(T\vert\theta) : \theta\in\Theta\right\}$ where $\theta\in\Theta$ 
is a real or integer-valued vector, is a family of probability density or mass functions $g(T\vert\theta)$. 
\emph{Parameter estimation} is then the problem of finding, among all the probability density/mass functions of $\mathcal{M}$, the $\hat{\theta}$ that most likely generated $T$. We estimate $\hat{\theta}$ with maximum likelihood estimation as follows: We seek the density/mass function that makes the observed term frequencies ``most likely'' \cite{myung2003tutorial} using a likelihood function, $\mathcal{L}(\theta\vert T)$, that specifies the likelihood of $\theta$ given $T$. $\hat{\theta}$ is then obtained by maximising the average
likelihood $
\hat{\theta}\! =\! \argmax_{\theta \in \Theta} \mathcal{L}(\theta \vert T)/\vert T\vert$ where $\mathcal{L}\left(\theta\vert T\right)\! =\! \sum_{i=1}^{\vert T\vert}\log g\left(f^i_{t,C}\vert\theta\right)$, and $f^i_{t,C}$ is the collection frequency of the $i$\textsuperscript{th} term $t\in T$. 

\subsection{Step 4: Best-fitting distribution}\label{sss:mc}
In step 4 of the ADR algorithm we select, from the set of candidate statistical distributions, the one that best quantifies the distribution of term frequencies for non-informative terms in $C$. We do this using Vuong's Closeness Test \cite{vuong1989likelihood} and Akaike's Information Criterion \cite{akaike1974}, both of which are among the least controversial and widely used statistical tests for model  selection \cite{clarke2003nonparametric}. Both tests are based on the Kullback--Leibler divergence and will favour the same model in the limit of large sample sizes, i.e.\ the model that minimises the information loss w.r.t.\ the unknown but ``true'' model. 

\subsection{Step 5: Distribution--adapted ranking}\label{ss:dfr}
Let $\hat{\mathcal{M}}$ be the best-fitting distribution to the collection term frequencies of the non-informative terms. Then, in step 5, we replace the 
distribution $P_1$ in Eqn.\ \ref{ch4:eqn:DFR} by $\hat{\mathcal{M}}$, yielding:
\begin{equation}\label{eqn:DFR-2}
R(q,d) = \textstyle\sum_{t\in q} f_{t,q}\cdot(-\log_2 \hat{\mathcal{M}})\cdot(1 - P_2)
\end{equation}
$\hat{\mathcal{M}}$ captures the same assumption as all DFR models: that non-informative terms are distributed in a certain way. However, whereas in DFR this assumption is not tested, ADR empirically validates the choice of $\hat{\mathcal{M}}$ per collection. 

\section{Evaluation}\label{s:eval}
We next evaluate the retrieval performance of our ADR models. As ADR can produce different ranking models for different datasets, we split this section into two parts. Section \ref{ss:deradr} presents our datasets and the ADR models instantiated for these datasets. Section \ref{ss:retexp} compares the retrieval performance of these ADR models against relevant baselines.


\subsection{Deriving ADR Models}\label{ss:deradr}

We use TREC disks 4 and 5 (TREC-d45) with queries 301-450 and 601-700 and ClueWeb cat.\ B.\ (CWEB09) with queries 1 - 200, minus query 20 for which no documents are judged relevant as datasets. All datasets are indexed without stop word removal and without stemming using Indri 5.10. For our ADR models, we must identify non-informative terms in each collection and the best fitting distribution of $f_{t,C}$ of the non-informative terms in the collection. 
We detect non-informative terms using SVM classification, as follows. Given an initial list of 40 informative and non-informative terms (manually compiled by three humans with complete agreement, see Table \ref{t:ini-terms}), we compute the term weight of these terms (using IDF \cite{sparck1972statistical}, $x^I$ \cite{2poisson}, residual-IDF (RIDF) \cite{church1999inverse}, gain \cite{papineni2001inverse} and $z$-measure \cite{harterp1} - see \cite{LiomaB09} for an overview of term weighting), and use these term weights as features to train a binary SVM classifier using 10-fold cross-validation on all feature combinations. \textcolor{black}{We use an SVM approach instead of any of the above term-weighting approaches, as the latter ones require (manual) setting some ad hoc threshold above/below which terms are considered informative/non-informative}. In contrast, the best classifier correctly classifies $\propto 86\%$ of the terms using RIDF and gain term weights as features. We use this classifier to classify each term in TREC-d45 and CWEB09 as informative or non-informative.

\begin{table}[!ht]
\centering
\scriptsize
\def\arraystretch{0.9}
\begin{tabular}{ll|ll}
\multicolumn{2}{c}{Informative terms} & \multicolumn{2}{c}{Non-informative terms}\\\hline
hypothermia	& intercepted & welcome & awesome \\
congolese	& furloughs & beginner  & jolly \\
outpaced	& randomization & spade & out\\
anthropocentric	& existentialist & delete & temp \\
iridescence	    & canvass & quit & feels \\
archdiocesan	& colonisation & least & loss \\
nonconformist	& airbrushed & silly  & clear \\
overclocking	& leviathans & cent & off \\
nominalization	& inflammation& jar & test \\
translucent	    & handmade & chat & fork \\
shortwave       & monasticism & yards & move \\
crystallography	& aperitif & fast & pair \\
expressionist	& pathologize & view & stop \\
cephalopod	    & abolishment & fold & colour \\
paparazzi	    & bookcase & roll & sit \\ 
presided	    & hydraulic & follow & back \\
beneficiary	    & vested & pen & hello \\
convection      & floods & day & stair \\
custodian       & chivalry & flash & best \\
populist	    & constrained & money & considerable\\\hline
\end{tabular}
\caption{Informative and non-informative terms.}
\label{t:ini-terms}
\end{table}

Next, using steps 3 -- 4 of the ADR algorithm we find that the best-fitting statistical model to the distribution of non-informative term frequencies in both datasets is the discrete Yule--Simon (YS) with parameter $p\!=\!1.804$ (TREC-d45) and $p\!=\!1.627$ (CWEB09).  
 The YS distribution is defined for $x\in\mathbb{Z}^+, x \geq 1$ \cite{simon1955class}:
\begin{equation}\label{eqn:ys_dist}
\mbox{P}\left(x\vert p\!+\!1\right) = (p+1)\cdot\frac{\Gamma(x)\cdot\Gamma\left(p+1\right)}{\Gamma(x + p + 1)}
\end{equation}
where $p\!>\!0$ is the distribution parameter and $\Gamma$ is the gamma function. Replacing $x$ with $\hat{f}_{t,d}$ (see Section \ref{s:dfr}), we obtain:
\begin{equation}\label{eqn:ys_model}
\mbox{P}^{\tiny\mbox{YS}}_1(\hat{f}_{t,d}\vert p\!+\!1) = (p+1)\cdot\frac{\Gamma(\hat{f}_{t,d})\cdot\Gamma(p + 1)}{\Gamma(\hat{f}_{t,d} + p + 1)}
\end{equation}
which, for a term $t$, returns the probability of having $\hat{f}_{t,d}$ occurrences of $t$. Identically to the original Poisson model \cite{Amati:2002:PMI:582415.582416}, Eqn.\ \ref{eqn:ys_model} is theoretically dubious as it is a discrete distribution whose input is a real number. Consequently, we use Lanczo's method to approximate the $\Gamma$ function identical to \cite{Amati:2002:PMI:582415.582416}. Plugging Eqn.\ \ref{eqn:ys_model} into Eqn.\ \ref{ch4:eqn:DFR} gives:
\begin{equation}\label{eqn:yuleadr}
R(q,d) = \textstyle\sum_{t\in q} f_{t,q}\cdot\left(-\log_2 \mbox{P}^{\tiny\mbox{YS}}_1\right)\cdot\left(1 - P_2\right)
\end{equation}
We refer to Eqn.\ \ref{eqn:yuleadr} as our YS ADR model, or YS for short.

\begin{figure}[!ht]
\centering
\subfloat[\small TREC Disks 4 \& 5 \label{f:idfscores}]{
    \includegraphics[width=\figwidth]{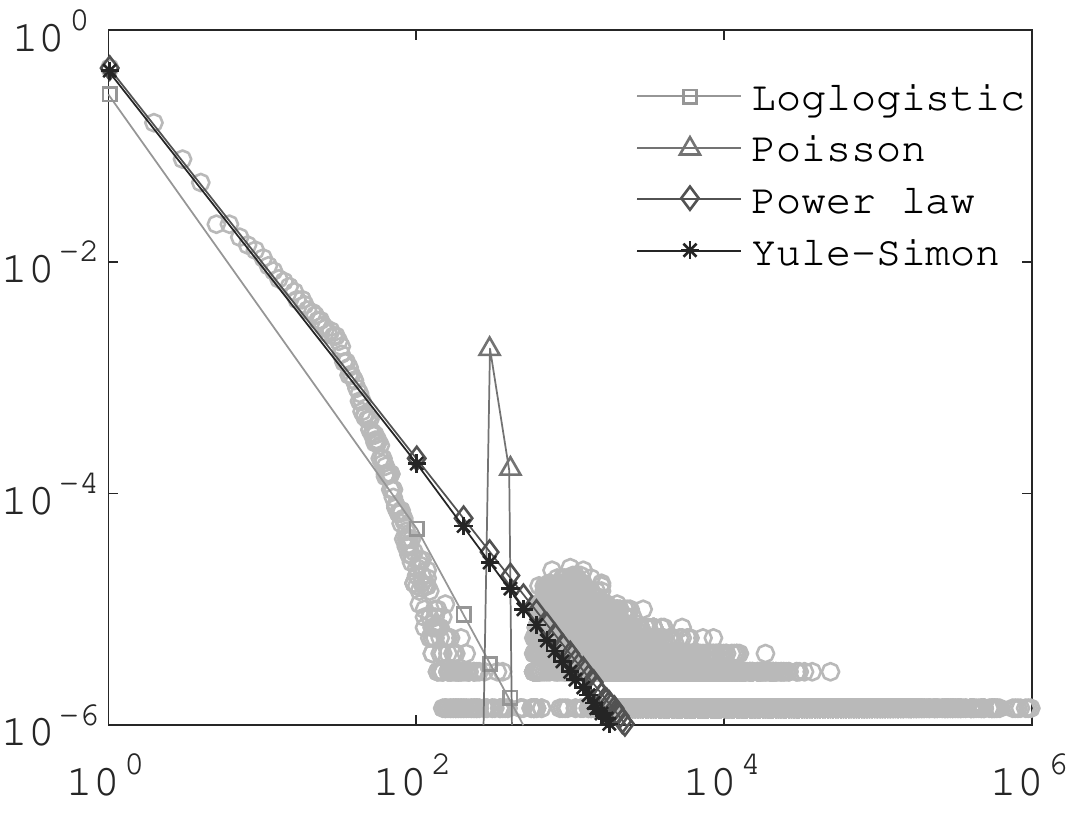}
}%
\subfloat[\small CWEB09 \label{f:ridfscores}]{
    \includegraphics[width=\figwidth]{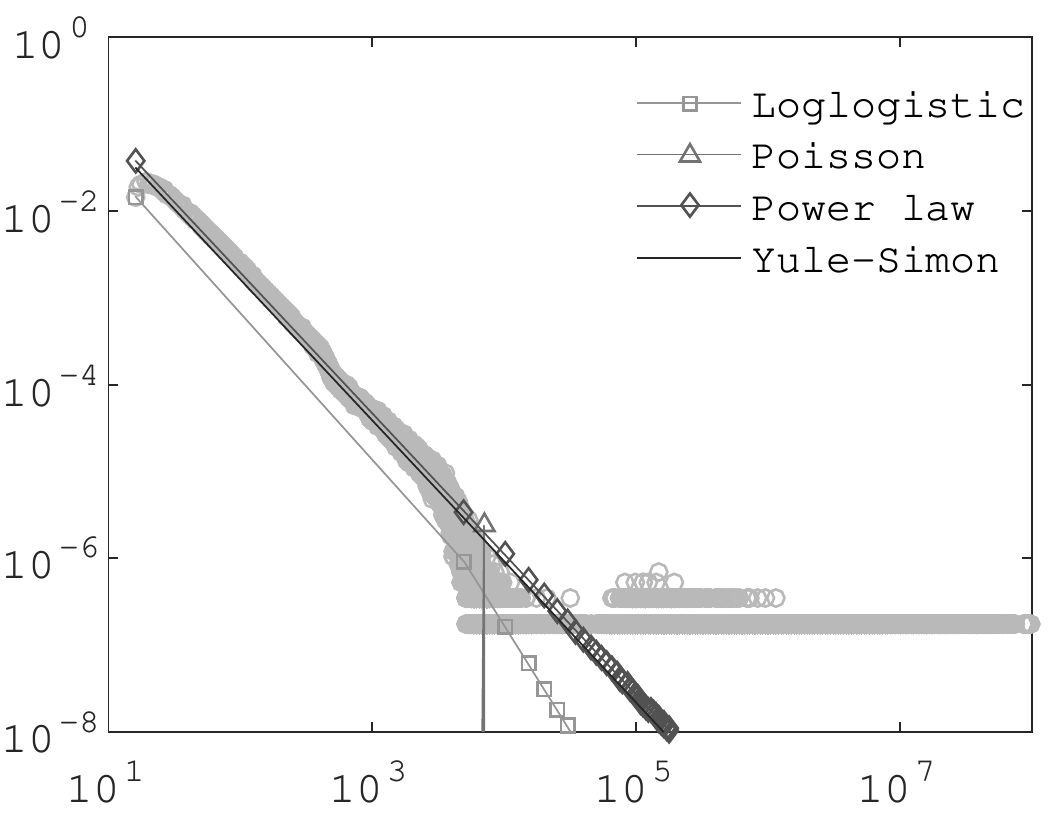}
}%
\caption{Collection term frequencies for non-informative terms (grey) in CWEB09, and TREC-d45. Superimposed on each figure are MLE fitted log-logistic, Poisson, power law and Yule--Simon distributions as references.} %
\label{f:fits}
\end{figure}

\subsection{Retrieval Experiments}\label{ss:retexp}

\noindent{\textbf{Baselines and Tuning.}} We compare our ADR ranking model (YS) to a Poisson (P), tf-idf (I$_n$), log-logistic (LL), smoothed power law (SPL) and query-likelihood unigram LM with Dirichlet smoothing (LMDir). P and $I_n$ are original DFR models \cite{Amati:2002:PMI:582415.582416}; LL \cite{DBLP:conf/ictir/ClinchantG09} and SPL \cite{Clinchant:2010:IMA:1835449.1835490} are information models that extend the original DFR models. All DFR/ADR models are postfixed with L2 meaning that Laplace and logarithmic term normalisation (see Section \ref{s:dfr}) are used. Following previous work \cite{Amati:2002:PMI:582415.582416,Clinchant:2010:IMA:1835449.1835490}, the parameter values of the distributions used in DFR/ADR (e.g. Poisoon or Yule--Simon) are induced from the collection statistics in two ways (we experiment with both): $T_{tc}\! =\! f_{t,C}/\vert C\vert$, or $T_{dc}\! =\! n_t/\vert C\vert$. E.g., PL2-$T_{tc}$ is a Poisson model using Laplace and logarithmic term normalisation, where Poisson's $\lambda\! =\! T_{tc} \! = \! f_{t,C}/\vert C\vert$ (see Eqn.\ \ref{eqn:dfr_poisson}), and YSL2-$T_{dc}$ is a Yule--Simon ADR model using Laplace and logarithmic term normalisation, where Yule--Simon's $p\! =\! T_{dc}\! =\!  n_t/\vert C\vert$ (see Eqn.\ \ref{eqn:ys_dist}). We retrieve the top-1000 documents according to each ranking model using title-only queries, and measure performance using P@10, Bpref, ERR@20, nDCG@10 and nDCG. All models are tuned using $3$-fold cross-validation and we report the average over all three test folds. We vary $c$ (the free parameter in the logarithmic term normalisation - see Section \ref{s:dfr}) in the range $\{0.5,1,2,4,6,8\}$ \cite{Clinchant:2010:IMA:1835449.1835490}, and $\mu$ of LMDir in the range $\{100,500,800,1000,2000,3000,4000,5000,8000,10000\}$.

\begin{table*}[!]
\centering
\small
\scalebox{0.999}{
\begin{tabular}{l | c c c c c | c c c c c |}
\hline
& \multicolumn{5}{c}{TREC disks 4\& 5}  & \multicolumn{5}{c}{ClueWeb09 cat.\ B.}\\
\hline
Model                 & nDCG & P@10  & Bpref &ERR@20 & nDCG@10  & nDCG & P@10  & Bpref &ERR@20 & nDCG@10\\
\hline
LMDir                  & .4643 & .3845 & .2239 & .1043 & .3968 &.2973 & .2586         & .2209 &  .0973 & .1769 \\
\hline
PL2-$T_{tc}$ \cite{Amati:2002:PMI:582415.582416}      & .2524$^*$ & .1273$^*$ & .1009$^*$ &  .0359$^*$ & .1332$^*$   &.1448$^*$ & .0712$^*$   & .1258$^*$          & .0211$^*$ & .0472$^*$\\
PL2-$T_{dc}$ \cite{Amati:2002:PMI:582415.582416}	& .2487$^*$ & .1217$^*$ & .0960$^*$ &  .0347$^*$ & .1273$^*$   &.1444$^*$ & .0709$^*$         & .1252$^*$          & .0314$^*$ & .0471$^*$ \\
I$_n$L2-$T_{tc}$ \cite{Amati:2002:PMI:582415.582416}   & .2917$^*$ & .1627$^*$ & .1114$^*$ &  .0478$^*$ & .1742$^*$   &.1596$^*$ & .0782$^*$         & .1405$^*$          & .0352$^*$ & .0511$^*$\\
I$_n$L2-$T_{dc}$ \cite{Amati:2002:PMI:582415.582416}  & .2818$^*$ & .1626$^*$ & .1088$^*$ &  .0481$^*$ & .1745$^*$   &.1596$^*$ & .0783$^*$         & .1407$^*$          & .0352$^*$ & .0512$^*$\\
\hline						  
LLL2-$T_{tc}$ \cite{DBLP:conf/ictir/ClinchantG09}  & \cellcolor{gray!25}.4812 & \cellcolor{gray!25}.4049 & \cellcolor{gray!25}.2341 &  \cellcolor{gray!25}.1072 & \cellcolor{gray!25}.4142   &\cellcolor{gray!25}.3184 & .2542         & \cellcolor{gray!25}.2349           & .0926 & .1706 \\
LLL2-$T_{dc}$ \cite{DBLP:conf/ictir/ClinchantG09} & \cellcolor{gray!25}.4810 & \cellcolor{gray!25}.3982 & \cellcolor{gray!25}.2329 &  \cellcolor{gray!25}.1069 & \cellcolor{gray!25}.4097 &\cellcolor{gray!25}.3180 & .2542         & \cellcolor{gray!25}.2349           & .0928 & .1707\\
SPLL2-$T_{tc}$ \cite{Clinchant:2010:IMA:1835449.1835490}   & \cellcolor{gray!25}.4863 & \cellcolor{gray!25}.4144 & \cellcolor{gray!25}.2375&  \cellcolor{gray!25}.1103 & \cellcolor{gray!25}.4276  &\cellcolor{gray!25}.3207 & .2529         & \cellcolor{gray!25}.2357          & .0945 & .1720  \\
SPLL2-$T_{dc}$ \cite{Clinchant:2010:IMA:1835449.1835490} & \textbf{\cellcolor{gray!25}.4876} & \cellcolor{gray!25}.4176&  \textbf{\cellcolor{gray!25}.2387}& \cellcolor{gray!25}.1107 & \cellcolor{gray!25}.4299    &\cellcolor{gray!25}.3224 & .2586         & \cellcolor{gray!25}.2370          & .0958 & .1752\\\hline
YSL2-$T_{tc}$ (ADR) & \cellcolor{gray!25}.4644 & \cellcolor{gray!25}.3982&  \cellcolor{gray!25}.2280& \cellcolor{gray!25}.1048&\cellcolor{gray!25}.4069 &\cellcolor{gray!25}.3197 & \cellcolor{gray!25}.2601         & \cellcolor{gray!25}.2359          & .0951 & .1752\\
YSL2-$T_{dc}$ (ADR)& \cellcolor{gray!25}.4860 & \cellcolor{gray!25}\textbf{.4182}&  \cellcolor{gray!25}.2381&  \cellcolor{gray!25}\textbf{.1113}&\cellcolor{gray!25}\textbf{.4312} &\textbf{\cellcolor{gray!25}.3240} & \textbf{\cellcolor{gray!25}.2666}         & \textbf{\cellcolor{gray!25}.2376}          & \textbf{\cellcolor{gray!25}.0985} & \textbf{\cellcolor{gray!25}.1810}\\\hline
\end{tabular}
}
\caption{Retrieval performance. Grey denotes larger than the LMDir baseline. Bold marks the best results. $^*$ marks statistically significant difference from the LMDir baseline using a $t$-test at the $.05\%$ level.} 
\label{t:trecbase}
\end{table*}

\begin{figure}
\centering
\includegraphics[width=0.8\columnwidth]{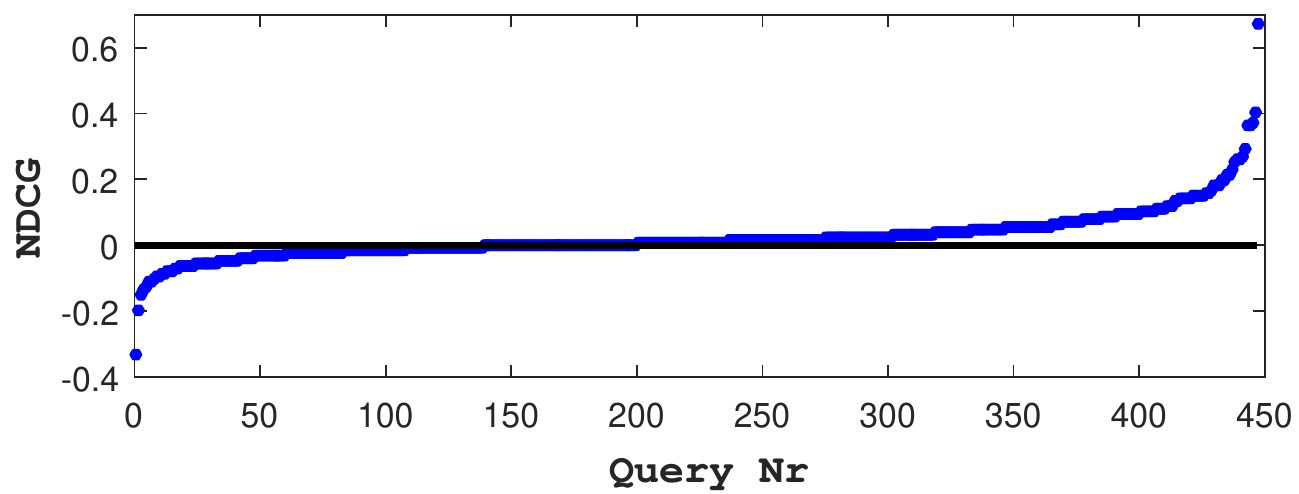}
\caption{Per query difference in nDCG between YSL2 and LMDir for all queries. Horizontal line indicates no difference. Points above zero favour YSL2 over LMDir.}
\label{fig:diff}
\end{figure}

\noindent{\textbf{Findings.}}\label{sss:results}
The results are shown in Table \ref{t:trecbase}. 
For CWEB09, our YSL2 models obtain the best performance at all times.
For TREC-d45, our YSL2 models obtain the best performance on early precision retrieval (P@10, ERR@20 and nDCG@10), while the SPLL2 models perform best on nDCG and Bpref (with YSL2 following closely). As Fig. \ref{f:fits} shows, the power law (used in SPLL2) and Yule--Simon (used in YSL2) have very similar fits, which explains their similar scores in Table \ref{t:trecbase}. In Fig. \ref{f:fits}  we also see that the log-logistic, power law and Yule--Simon all approximate the head of the distribution (collection term frequencies up to $\approx 100$), though all fail to model the ``noisy'' tail (collection term frequencies $> 1000$) well. 
In contrast, the Poisson distribution fails to model the empirical data in any range, possibly explaining the overall low retrieval performance in Table \ref{t:trecbase}. That the Yule--Simon distribution gives the highest retrieval performance indicates the merit of using the best-fitting distribution for DFR ranking models. Fig.\ \ref{fig:diff} shows the per-query difference in nDCG between the YSL2 and LMDir baseline for all queries. We see that the number of queries that benefit from YSL2 is substantially higher (270 queries) than for the LMDir (163 queries), confirming the scores in Table \ref{t:trecbase}.


\section{Related Work}
\label{s:rw}
Our work can be seen as a refinement of DFR ranking models.
Several other approaches attempt such refinements. For example, Clinchant and Gaussier \cite{DBLP:conf/ictir/ClinchantG09} formulated heuristic retrieval constraints to assess the validity of DFR ranking models, and found that several DFR ranking models do not conform to these heuristics. On this basis, a simplified DFR model based on a log-logistic (LL) statistical distribution was proposed, which adheres to all heuristic retrieval constraints. Experimental evaluation showed that the LL model improved MAP on several collections. In a follow-up study, Clinchant and Gaussier \cite{Clinchant:2010:IMA:1835449.1835490} introduced \emph{information models} which are simplified DFR models that do not rely on so-called after-effect normalisation. They instantiated the LL and a ``smoothed power law'' (SPL) ranking model, and showed that both models tend to outperform a LM, BM25, but not two original DFR models on smaller TREC and CLEF collections. Closest to ours is the work by Hui et al.\ \cite{Hui:2011:RWU:2063576.2063595} who developed DFR extensions where the distributions of $f_{t,C}$ are modelled using multiple 
statistical distributions. However, (i) no justification or empirical validation of the choice of statistical distributions used was given, and (ii) the distribution of all terms, rather than only the non-informative terms, was used. Finally, Hui et al's models also removed e.g.\ collection-wide term frequencies from the ranking component, hence effectively removing the notion of divergence from the DFR framework.

\section{Conclusion}
Divergence From Randomness (DFR) ranking models assume that informative terms in a collection are distributed differently than non-informative terms. Different DFR models are produced depending on the statistical model (e.g.\ Poisson, geometric) used to quantify the distribution of non-informative terms, where an informative term is detected by measuring the divergence of its distribution from the distribution of non-informative terms. However, there is little evidence that the distributions used in DFR actually fit the the distribution of non-informative terms. To address this, we presented a novel DFR extension called Adaptive Distributional Ranking (ADR), which adapts the ranking computation to the statistics of the specific collection being processed each time. Our ADR models first determine the best-fitting distribution of non-informative terms, and then integrate this distribution into ranking. Experiments on TREC data showed that our ADR models outperformed DFR models (and their extensions), and achieved performance comparable to a query likelihood LM. In future work we plan to experiment with additional collections. We will also study automatically deriving ad hoc distributions suited for the collection data in ADR instead of selecting among a list of standard distributions.

\section{Acknowledgments}
 Work partially funded by C. Lioma's FREJA research excellence fellowship (grant no. 790095).

\bibliographystyle{abbrv}
{\normalsize
\bibliography{cikm-red}  
}
\end{document}